\documentclass[aps,prd,twocolumn]{revtex4-2}
\usepackage{amsmath, amssymb, physics, graphicx}
\usepackage[margin=0.7in]{geometry}
\usepackage{tabularx}
\usepackage[table,dvipsnames]{xcolor}
\usepackage{array}
\usepackage{mathtools} 
\usepackage{tikz}
\usetikzlibrary{arrows.meta,calc}
\usepackage{bm}  

\usepackage[normalem]{ulem}   

\usepackage[colorlinks=true, linkcolor=blue, citecolor=red, urlcolor=blue]{hyperref}
\newcolumntype{Y}{>{\centering\arraybackslash}X}

\newcommand{\jb}{J_{\textrm{b}}}
\newcommand{\yb}{y_{\textrm{b}}}

\newcommand{\basinfig}{%
\begin{figure}[tb]
  \centering
  \begin{tikzpicture}
    \node[anchor=south west, inner sep=0] (img) at (0,0) {
      \includegraphics[width=0.42\linewidth]{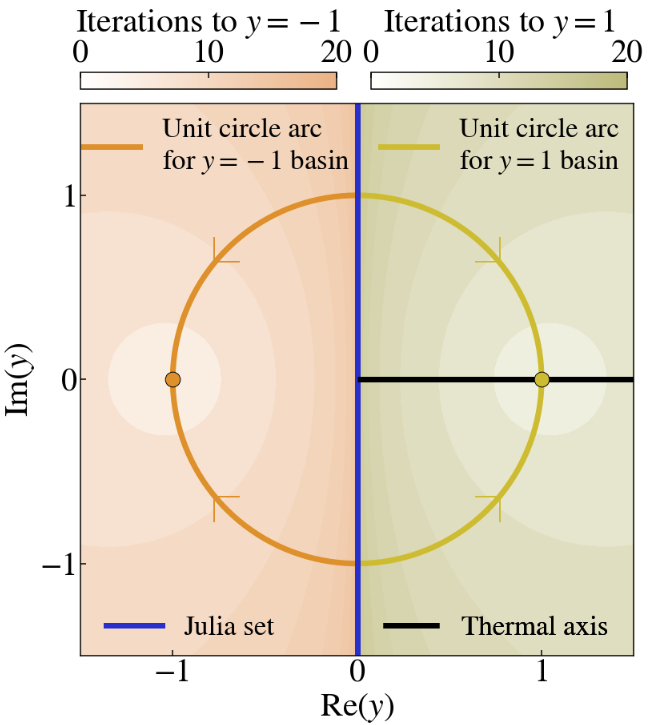}
      \includegraphics[width=0.55\linewidth]{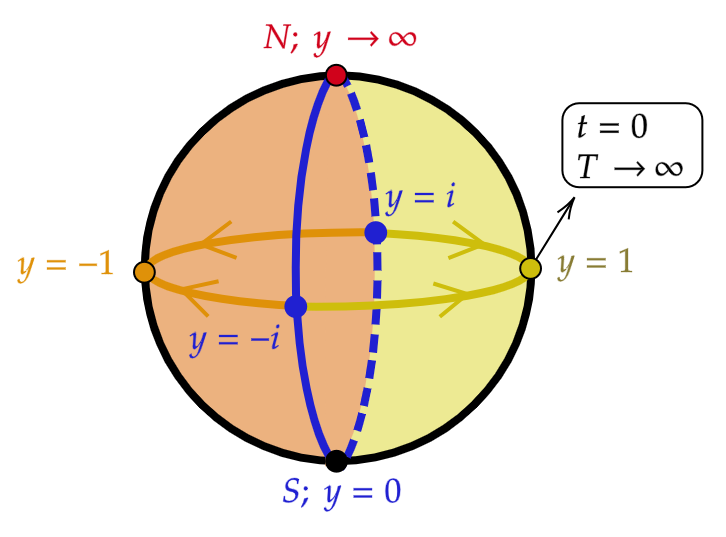}
  };
    \begin{scope}[x={(img.south east)}, y={(img.north west)}]
      \node at (0.45,0.12) { \textbf{(a)}};
      \node at (0.85,0.12) { \textbf{(b)}};
    \end{scope}
  \end{tikzpicture} 
  \caption{(a) The complex $y$-plane.  The positive x-axis (Re~$y>0$)
    describes the thermal problem while the quantum evolution is along
    the unit circle ($|y|=1$). The basins of attraction for the RG map
    in the complex y-plane are colour-coded. (b) Representation on the
    Riemann sphere for the extended complex plane. The unit circle for
    quantum dynamics is the equator. The unstable fixed point at
    infinity is the North pole, N, (red dot), the South pole (S) being
    the origin. The Julia set is the dashed blue circle through the
    two poles with the DQPT points marked as blue dots.}
\label{fig:basins}
\end{figure}
}%

\newcommand{\rgpicture}{%
  \begin{figure}[tb]
    {\centering
    \includegraphics[width=\linewidth,trim=0pt 0pt 0pt
    0pt,clip]{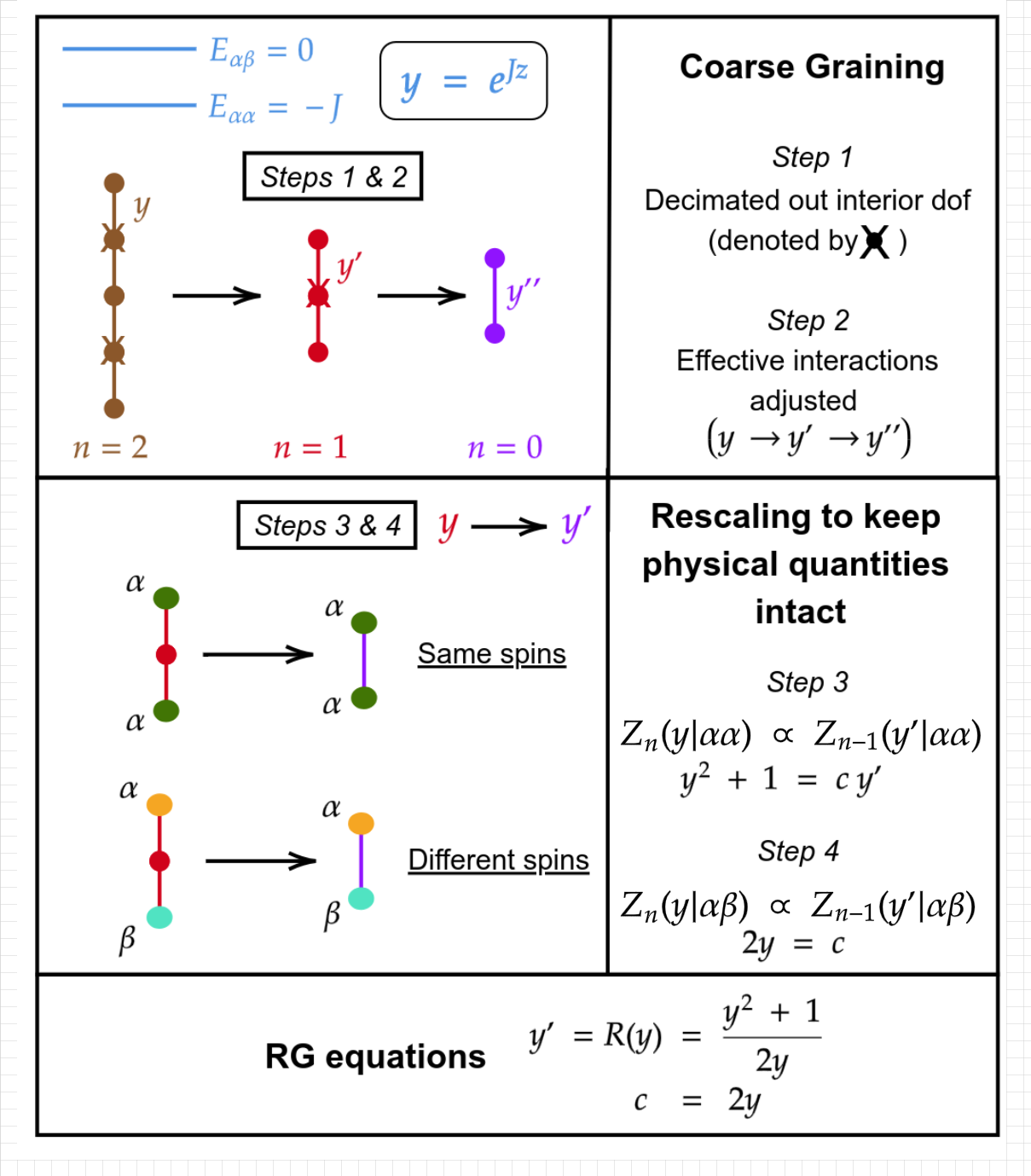}
    }
    \caption{Schematic representation of the RG procedure: (1)
      coarse-graining by eliminating interior degrees of freedom, (2)
      adjustment of effective interactions, and (3--4) rescaling with
      renormalized parameters to preserve macroscopic quantities such
      as the partition function $Z_n(y|\alpha\beta)$, where $\alpha,
      \beta$ denote the states of the boundary spins at the nth
      generation. }
    \label{fig:RGsteps}
  \end{figure}
}%

\newcommand{\trianglefig}{%
\begin{figure}
  \centering
  \includegraphics[width=0.75\linewidth]{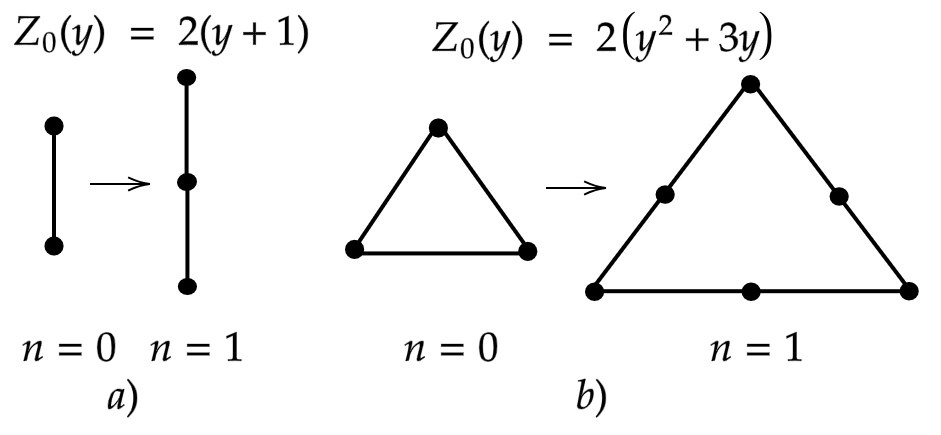}
  \caption{(a) One bond as the starting unit ($n=0$) for an open
    chain.  (b) A triangle as the basic unit ($n=0$) for the periodic
    chain. The partition functions for the smallest structures are
    shown above the respective figures.}
  \label{fig:triangle}
\end{figure}
}%

\newcommand{\dqptplot}{%
  \begin{figure}[htbp]
    \centering
    \includegraphics[width=\linewidth]{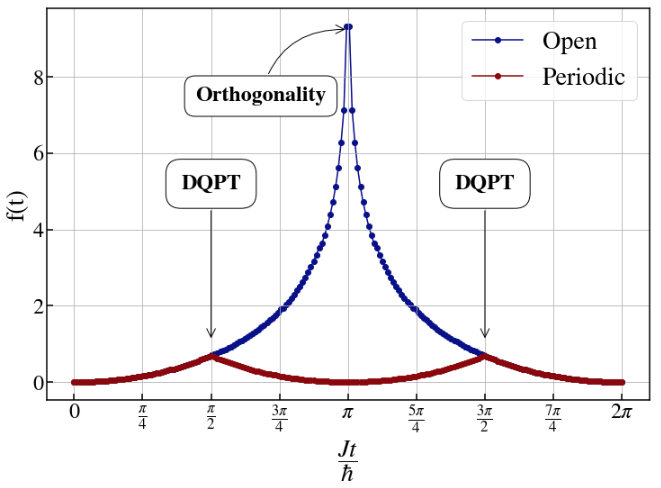}
    \caption{The free energy per spin $f(t)$ for the one-dimensional
      transverse-field Ising model \cite{dataset}.  The periodic chain
      (red) shows nonanalytic cusps at critical times.  The value of
      $f(t)$ at the transition point is $(\ln 2)/2$.  The free energy
      $f(t)$ for the open chain (blue) remains smooth around
      transition points. The log divergence corresponds to the
      orthogonal catastrophe.}
    \label{fig:open-vs-periodic}
  \end{figure}
}%

\newcommand{\ringopen}{%
  \begin{figure}[tb]
    {\centering
    \includegraphics[width=0.75\linewidth]{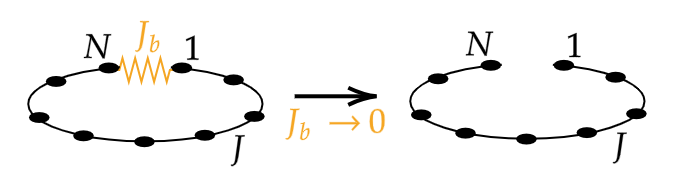}}
  \caption{Two geometries: (a) a periodic chain, (b) an open chain.
    The link connecting site 1 to site $N$ has a coupling $J_b$ while
    others have $J$. The open chain geometry is achieved by cutting
    the boundary bond or by setting $J_b=0$.}
    \label{fig:ring}
  \end{figure}
}%

\newcommand{\tmzeroonetwo}{%
\begin{figure*}[tb]
    {\centering
      \begin{tikzpicture}
    \node[anchor=south west, inner sep=0] (img) at (0,0) 
    {\includegraphics[width=0.3\linewidth,trim=0pt 0pt 0pt 0pt,clip]{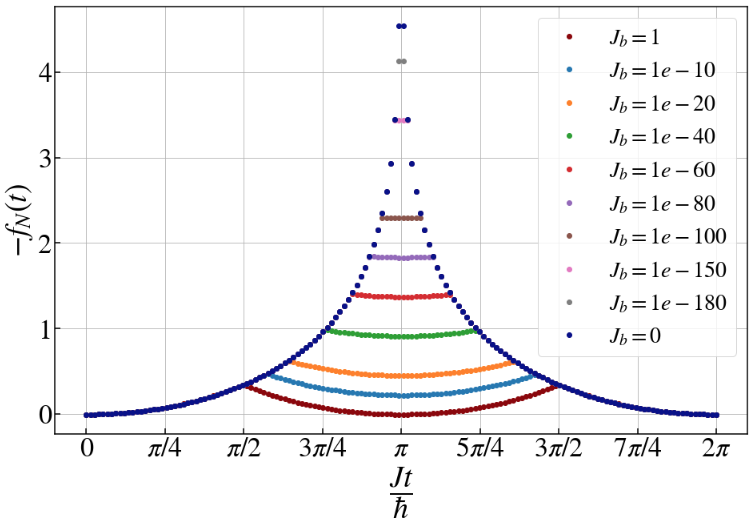}
    \includegraphics[width=0.3\linewidth,trim=0pt 0pt 0pt 0pt,clip]{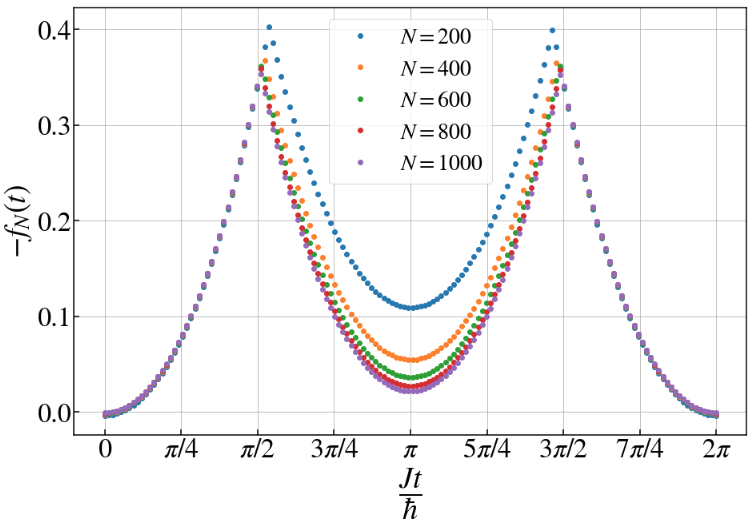}
    \includegraphics[width=0.3\linewidth,trim=0pt 0pt 0pt 0pt,clip]{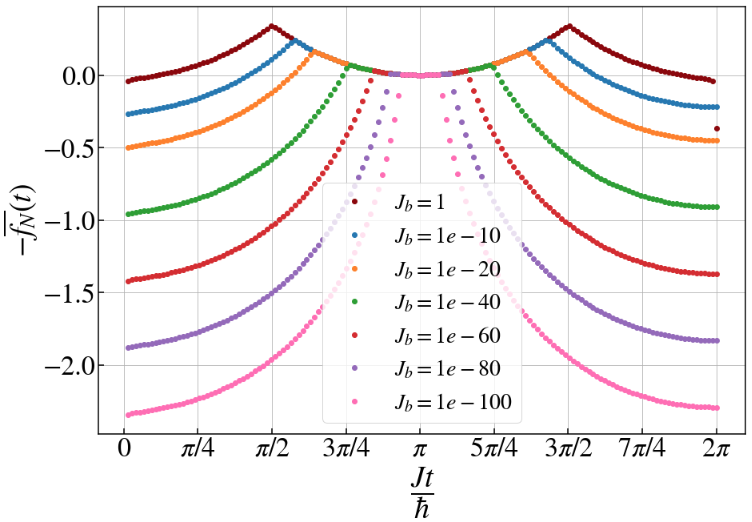}};
    \begin{scope}[x={(img.south east)}, y={(img.north west)}]
      \node at (0.17,1.05) { \textbf{(a)}};
      \node at (0.5,1.05) { \textbf{(b)}};
      \node at (0.86,1.05) { \textbf{(c)}};
    \end{scope}
  \end{tikzpicture} 
}%
\caption{(a) The free energy $-f_N(t)$ (Eq. \eqref{eq:frecur})for
  various boundary couplings $\jb$ for $N=100$.  (b) The free energy
  $-f_N(t)$ for a fixed boundary coupling $\jb=10^{-10}$ for various
  $N$.  (c) $-\overline{f}_N(t)$, the scaled free energy, showing data
  collapse for the intermediate phase.  See Ref. \cite{dataset}.}
    \label{fig:TM_012}
\end{figure*}
}%

\begin{document}
\title{Julia Set in Quantum Evolution: The case of Dynamical Quantum Phase Transitions}

\author{Manmeet Kaur } 
\email{manmeetkaur933\string@gmail.com }
\author{Somendra M. Bhattacharjee }
\email{ somendra.bhattacharjee\string@ashoka.edu.in}
\affiliation{ Department of Physics, Ashoka University, Sonepat,
  Haryana - 131029, India} 

\begin{abstract}
  Dynamical quantum phase transitions (DQPTs) are a class of
  non-equilibrium phase transitions that occur in many-body quantum
  systems during real-time evolution, rather than through parameter
  tuning as in conventional phase transitions.  This paper presents an
  exact analytical approach to studying DQPTs by combining complex
  dynamics with the real-space renormalization group (RG).  RG
  transformations are interpreted as iterated maps on the complex
  plane, establishing a connection between DQPTs and the Julia set,
  the boundary separating the basins of attraction of the stable fixed
  points.  This framework is applied to a quantum quench in the
  one-dimensional transverse field Ising model, where we examine the
  sensitivity of DQPTs to variations in boundary conditions. We show
  that altering the topology of the spin chain can suppress DQPTs and
  provide a qualitative explanation based on quantum speed limits.
\end{abstract}

\maketitle

\section{Introduction}
Phase transitions signify qualitative changes in the large-scale
behavior of physical systems, typically indicated by nonanalyticities
in thermodynamic observables when external parameters, such as
temperature, pressure, or applied fields, are varied. Classic examples
include the onset of ferromagnetism or the condensation of a gas into
a liquid.  In quantum systems at zero temperature, similar
transitions, known as quantum phase transitions, can occur due to
quantum fluctuations when a parameter in the Hamiltonian is tuned.
These quantum phase transitions (QPTs) involve distinct phases with
differing ground state characteristics \cite{suzuki,vojta}.  More
recently, it has been recognized that isolated many-body quantum
systems can show nonanalytic behaviour in time during their unitary
evolution following a sudden quench.  These are called dynamical
quantum phase transitions (DQPTs) and are signaled by singularities in
the rate function, which plays a role analogous to the free energy
density in equilibrium thermodynamics 
\cite{heylreview,heylprl,zvyagin,amina,smbpotts}.  Unlike conventional
transitions, where time merely parametrizes dynamics, DQPTs exhibit
transitions that recur periodically in real time.  This perspective
challenges the traditional role of time in quantum and statistical
mechanics and highlights its potential as a control variable in
nonequilibrium phenomena.
 
In standard QPTs, changing a parameter in the Hamiltonian modifies the
energy spectrum, and the transition is typically associated with a
vanishing energy gap in continuous cases, or a level crossing between
competing ground states in first-order transitions \cite{vojta}. In
contrast, DQPTs occur under a fixed Hamiltonian when quenched from a
non-eigenstate.  Since the Hamiltonian is fixed during the evolution,
the energy spectrum is also fixed. The transition is instead caused by
a modification of the complex amplitudes, especially the phase
factors, of the energy eigenstates, in the thermodynamic limit.

Quench dynamics has emerged as a central theme in contemporary quantum
many-body systems, with relevance spanning from fundamental questions,
such as thermalization \cite{quench}, entanglement growth
\cite{entang}, and non-equilibrium dynamics
\cite{fischer,krish,paraj1,paraj2,nechaev}, to practical applications
in quantum computation and information processing
\cite{vajna,yuanfidelity}.  Among various phenomena in this context,
DQPTs have attracted significant attention
\cite{heyl2,expt2,experiment,banuls,halimeh2023,Andras,pozsgay,maslowski,johann,jafari2024,jafari2025,yu2021,uma,zhou,mishra,niccolo,smbpotts}.
Initially identified in the transverse field Ising model
\cite{heylprl}, these transitions have since been investigated in a
wide range of settings.  Despite this progress, a key open question
remains: To what extent do the critical behaviors observed in DQPTs
reflect analogues of equilibrium thermal phase transitions, and under
what conditions might genuinely new phases or critical phenomena
emerge?  Addressing this question is essential for advancing our
understanding of non-equilibrium quantum matter and its possible
classification.

Recently, real space renormalization group (RG) methods have been
employed to study DQPTs exactly in a variety of scale-invariant
lattices across different spatial dimensions \cite{amina,smbpotts}.
The RG treatment of these transitions involves iterating the
transformation in a suitably chosen complex parameter plane. This
approach draws upon techniques from complex dynamical systems,
particularly the theory of iterated maps of complex variables
\cite{milnor,beardon,carleson}, and has been demonstrated for models
such as the three-state quantum Potts model \cite{smbpotts}.

In this paper, we apply this complex RG framework to the
one-dimensional transverse-field Ising model (TFIM), with particular
emphasis on the emergence and role of the Julia set in characterizing
DQPTs. In standard RG analysis, stable fixed points correspond to
distinct phases, while unstable fixed points indicate criticality. In
the complex plane, however, the picture becomes richer: each stable
fixed point possesses a basin of attraction, and the boundary between
these basins is a dense set, often a fractal, known as the Julia set
\cite{milnor,beardon,carleson}.  While the unstable fixed points are
embedded within the Julia set, it is the global structure of this set
that captures the essential features of these transitions in the
quantum system.

The remainder of this paper is organized as follows.  The formalism of
DQPT and the connection between the Loschmidt amplitude and the
partition function is shown in Sec. \ref{sec:formalism}.  The analytic
continuation helps in identifying a free energy like quantity (the
rate function in large deviation theory) as done in Sec.
\ref{sec:part-funct-large}. The model (TFIM), its symmetry,
symmetry-breaking, and the associated quench protocol are discussed in
Sec. \ref{sec:protocol-dqpt-tfim}. This section introduces the
variable $y$ which, as a complex variable, helps in the analytic
continuation of the partition function and the Loschmidt amplitude.
The basic RG transformation in the complex plane is done in Sec.
\ref{sec:rg} which also contains some of the features of complex
dynamics and the Julia set.  The Julia set for the 1-D TFIM in the
complex-$y$ plane and the connection to the zeros of the partition
function is pointed out in Sec. \ref{sec:zeros-1-d}.  This connection
is feasible because the RG transformation is based on the partition
function and its invariance under RG. The DQPT and the surprising
features of the one dimensional problem are discussed in Sec.
\ref{sec:basins-attr-dqpt}.  The role of boundary conditions, periodic
chain vs open chain, is analyzed in Sec. \ref{sec:interm-phase-bound}.
The sensitivity to boundary conditions \cite{amina} is unique to DQPT
and is not generally seen in thermal phase transitions.  A simple
argument based on quantum speed limits to explain the difference is
also given there.  The paper concludes in Sec. \ref{sec:conclusion}.
Appendix \ref{sec:rg-relations} contains further details on the RG
procedure.  Features of complex dynamics and the Julia set are
explored in
Appendices\ref{sec:degree-rational-map}-\ref{sec:julia-set-rg}.  In
particular, Appendix \ref{sec:why-not-fractal} discusses the
conditions because of which the Julia set for the one dimensional
chain is an algebraic curve.  This is not the case for higher
dimensional models \cite{amina,smbpotts} which exhibit fractal Julia
sets.  These higher dimensional cases will be discussed elsewhere.
Appendix \ref{sec:density-zeros} discusses the density of zeros of the
partition function.

\section{The Formalism of Dynamical Quantum Phase Transitions}
\label{sec:formalism}

We consider an \(N\)-particle quantum system described by a
Hamiltonian \(H\) with a complete set of orthonormal eigenstates
\(\{\ket{n}\}\) and corresponding eigenvalues \(\{E_n\}\).  The system
is initially prepared in the pure state
\begin{equation}
  \ket{\psi_0} = \sum_{n=1}^{\mathcal{W}} c_n(0) \ket{n},\quad
  \left(\sum_{n=1}^{{\cal{W}}} |c_n(0)|^2=1\right), 
    \label{eq:initial_state_general}
\end{equation}
where \(\mathcal{W}\) is the total number of eigenstates.  We choose
this initial state to \emph{not} be an eigenstate of \(H\), ensuring
that the subsequent unitary dynamics is nontrivial.  The time
evolution is given by (we set $\hbar = 1$),
\begin{equation}
  \ket{\psi_t} = e^{-i H t} \ket{\psi_0}= \sum_{n=1}^{\mathcal{W}} c_n(t) \ket{n},
    \label{eq:psi_t_evolution}
\end{equation}
so that $c_n(t)=c_n(0) e^{-itE_n}$.  A central quantity for probing
this evolution is the \textit{Loschmidt amplitude}
\cite{heylreview,heylprl,zvyagin}
\begin{equation}
    L(t) = \braket{\psi_0}{\psi_t},
    \label{eq:loschmidt_amp}
\end{equation}
which encodes the dynamics of a state in an exponentially large
Hilbert space into a single complex number.

The associated {Loschmidt echo} or return probability \cite{fidel} is
\begin{equation}
    P(t) = |L(t)|^2,
    \label{eq:loschmidt_echo}
\end{equation}
describing the probability of finding the system in its initial state
at time $t>0$.

We consider an initial state given in
Eq.~\eqref{eq:initial_state_general} as an equal superposition over
all \(\mathcal{W}\) eigenstates (similar to the initial states of
Refs. \cite{heylprl,heyl2})
\begin{equation}
    c_n(0) = \frac{1}{\sqrt{\mathcal{W}}}, \quad \forall n,
    \label{eq:equal_superposition}
\end{equation}
which uniformly explores the full spectrum, much like an
infinite-temperature state in statistical physics.  For this state,
the Loschmodth echo can be decomposed as
\begin{equation}
  \label{eq:3}
  P(t)= \frac{1}{\mathcal{W}} + \frac{2}{\mathcal{W}} \sum_{n>m}
  \cos\!\left[(E_n - E_m)t \right],
\end{equation}
which respects the expected bounds, $0\le P(t)\le 1$.  
As $|c_n(t)|^2 = |c_n(0)|^2$, for all $t$, the time dependence of $L(t)$ comes
entirely from the \textit{interference of evolving states}.  Different
eigenstates accumulate different phases, and their superposition can
strongly alter the system's macroscopic properties leading to DQPT.

\subsection{Partition function and large deviations}
\label{sec:part-funct-large}

For the state of Eq. \eqref{eq:equal_superposition}, the Loschmidt
amplitude, Eq. (\ref{eq:loschmidt_amp}), simplifies to
\begin{equation}
    L(t) = \frac{1}{\mathcal{W}} \sum_{n=1}^{\mathcal{W}} e^{-i E_n t},
    \label{eq:L_equal_superposition}
\end{equation}
closely resembling the canonical partition function of the considered
quantum system,
\begin{equation}
    Z(\beta) = \sum_{n} e^{-\beta E_n},
    \label{eq:partition_function}
\end{equation}
where $\beta = 1/T$ denotes the inverse temperature (we set
$k_{\text{B}} = 1$ throughout).  This resemblance is made precise by
an analytic continuation via the generalized function
\begin{equation}
    L(z) = \frac{1}{\mathcal{W}} \sum_{n=1}^{\mathcal{W}} e^{-z E_n}, 
    \label{eq:Lz_def}
\end{equation}
with $z \in \mathbb{C}$, which bridges the equilibrium and dynamical
descriptions, as
\begin{equation}
    L(z) =
    \begin{cases}
        \tfrac{1}{\mathcal{W}} Z(\beta), & z = \beta \in \mathbb{R}, \\[6pt]
        L(t), & z = i t,\ \ t \in \mathbb{R}.
    \end{cases}
    \label{eq:Lz_connection}
\end{equation}
Thus, the Loschmidt amplitude is understood as an analytic
continuation of the partition function as $\beta \to it$ in the
$(\beta,t)$ complex plane.

In the thermodynamic limit, the Loschmidt echo $P(t)$,
Eq.~\eqref{eq:loschmidt_echo}, acquires a large-deviation form
\cite{largedev}
\begin{equation}
    P(t) \sim e^{-2N f(t)},
    \label{eq:echo_large_dev}
\end{equation}
with the rate function defined as
\begin{equation}
    f(t) = - \Re \lim_{N \to \infty} \frac{1}{N} \log L(t),
    \label{eq:rate_function_def}
\end{equation}
where $N$ is the number of degrees of freedom.  Because of the
analytic continuation, this function behaves as the \textit{dynamical
  free-energy density}, or free energy in short, for our problem.  In
the thermodynamic limit, DQPT occurs when $f(t)$ becomes non-analytic
in $t$, signalling a reorganization of the microscopic phase-coherence
structure that produces a singularity in the macroscopic rate function
\cite{heylprl,fischer}.

The Loschmidt amplitude, $L(t)$, quantifies the overlap between the
initial quantum state and its time-evolved counterpart.  A vanishing
amplitude $L(t)=0$ indicates complete orthogonality between the two
states, signaling a total loss of memory of the initial configuration.
A prototypical example is when all spins, initially aligned along the
$+x$-direction, evolve into a configuration aligned along $-x$. While
such states may appear similar in the classical sense as vectors, they
are orthogonal in the Hilbert space. This loss of memory of the
initial state is referred to as the orthogonality catastrophe, often
accompanied by a logarithmic divergence in the rate function $f(t)$.
In many cases, however, this catastrophe is preempted by a DQPT,
whereby the system transitions into a macroscopically distinct state.
Both the appearance of this new state and the associated
non-analyticity are strictly defined only in the $N\to\infty$ limit,
highlighting their nature as collective, emergent phenomena in
nonequilibrium dynamics.

\section{The protocol for DQPT in TFIM}
\label{sec:protocol-dqpt-tfim}

\subsection{Transverse-Field Ising Model}
\label{sec:tfim}
We consider the transverse-field Ising model (TFIM), which serves as
the central model in this work. The Hamiltonian is given by
\cite{suzuki}
\begin{equation}
    H = H_J + H_F,
    \label{eq:TFIM_H}
\end{equation}
with the interaction and field contributions defined as
\begin{equation}
    H_J = -\frac{J}{2}\sum_{i}\big(1+\sigma_i^z\sigma_{i+1}^z\big),
   \quad  H_F = -\Gamma \sum_{i=1}^{N}\sigma_i^x,
    \label{eq:ising_field}
\end{equation}
where $\sigma_i^x, \sigma_i^z$ are the Pauli spin operators at site
$i$.  Here, $J>0$ is the ferromagnetic coupling and $\Gamma$ is the
transverse field strength. The sum in $H_J$ runs over the sites of a
one-dimensional chain; for open boundary conditions the sum is from
$i=1$ to $N-1$, while for periodic boundary conditions, it runs from
$i=1$ to $N$ with $\sigma_{N+1}^z \equiv \sigma_1^z$. We consider both
boundary conditions in this study.

The TFIM Hamiltonian is invariant under a global spin flip
transformation implemented by ${\cal{P}}=\prod_i \sigma_i^x$ under
which the Pauli matrices transform as
\begin{equation}
  \label{eq:10}
  {\cal{P}} \sigma_i^z {\cal{P}}^{-1} = -\sigma_i^z, {\text{ and }}
  {\cal{P}} \sigma_i^x {\cal{P}}^{-1} = \sigma_i^x,
\end{equation}
so that $[H,{\cal{P}}]=0$, reflecting a global $\bm{Z}_2$ symmetry \cite{suzuki}.
  
The interaction term \(H_J\) favors parallel alignment of neighbouring
spins along the $z$-axis, while the transverse field term \(H_F\)
promotes quantum fluctuations that tend to disrupt this ordering. The
competition between these two terms is well known to drive a
zero-temperature quantum phase transition that spontaneously breaks
the global ${\bm{Z}}_2$ symmetry.  In the limit $\Gamma\to\infty$, the
ground state is unique, with all spins aligned along the $+x$
direction, while for $\Gamma\to 0$, the system enters a
twofold-degenerate broken-symmetry phase, with spins aligned either
all up or all down along $z$.  The QPT \cite{suzuki} corresponds to
the transition between these two regimes at $\Gamma=J/2$.  This
underlying structure motivates the TFIM as a minimal framework for
studying quantum quenches, particularly those involving a sudden
change of the transverse field from $\Gamma\to\infty$ to $\Gamma\to
0$.

\subsection{The Quench Protocol}
\label{sec:quench}
\subsubsection{The protocol}
\label{sec:protocol}

For a sudden quench of the transverse field from $\Gamma\to\infty$ to
$\Gamma=0$, the system evolves unitarily under $H_J$.  Such a quench
has played a central role in identifying DQPTs in the transverse-field
Ising model (TFIM) \cite{heylreview,heylprl,zvyagin}. In this work, we
analyze the same quench using a real-space renormalization-group
framework formulated as an iterated map in the complex plane of the
Boltzmann factor, thereby linking the Julia set of the RG map with
both the thermal phase transition and the ensuing quantum dynamics.
This provides a complementary perspective on DQPTs and suggests
natural extensions to higher dimensions and other quantum models.

For $\Gamma \to \infty$, the initial state is prepared in the product
state
\begin{equation}
  \label{eq:14}
  \ket{\psi_0} = \bigotimes_{j=1}^{N} \ket{\rightarrow}_j,  
\end{equation}
where $\ket{\rightarrow}_j$ is the eigenstate of $\sigma_j^x$ with
eigenvalue $+1$.  Expressed in the $\sigma^z$ basis as
  $\ket{\rightarrow}_j=\frac{1}{\sqrt{2}}
  \left(\ket{\uparrow}_j+\ket{\downarrow}_j\right),$ 
where $\ket{\uparrow},\ket{\downarrow}$ denote spins in the z-direction.
Upon unfolding the factors, $\ket{\psi_0}$ can be written as in Eqs.
(\ref{eq:initial_state_general}) and (\ref{eq:equal_superposition}) as
\begin{equation}
  \label{eq:5}
   \ket{\psi_0} = \frac{1}{\sqrt{\cal{W}}}  \sum_{n=1}^{\mathcal{W}}
   \ket{n},\quad ({\cal{W}}=2^N),
\end{equation}
where $\ket{n}$ are the eigenstates of $H_J$, 
$H_J \ket{n}=E_n \ket{n}$, with
\begin{equation}
  \label{eq:11}
  E_n=-J\sum_k \delta(s_k,s_{k+1}),
\end{equation}
$s_k=\pm 1$ denoting the spin orientation (in the z-direction) of the
spin at site $k$.  The eigenstates consist of all possible
arrangements of the $\pm 1$ states along the chain. Here $\delta(p,q)$
is the Kroneker-$\delta$ which is 1 if $p=q$, 0 otherwise.

\subsubsection{The choice of variable and the complex plane}
\label{sec:choice-vari-compl}

\basinfig

In the representation of Eq. (\ref{eq:11}), two parallel neighboring
spins contribute an energy of $-J$, while anti-parallel spins
contribute zero, so that the energy gap is $J$.  We, therefore,
introduce the parameter
\begin{equation}
    y = e^{z J},\quad(z\in {\cal{C}})
    \label{eq:boltzmann_y}
\end{equation}
which, for real $z=\beta$, represents the relative Boltzmann weight
associated with the ordered pair, while for purely imaginary $z=it$,
it describes the dynamical phase under unitary time evolution.
Hereafter, all quantities will be expressed in terms of $y$ in Eq.
(\ref{eq:boltzmann_y}), unless stated otherwise.  In the complex-$y$
plane, Fig. \ref{fig:basins}, the thermal partition function is
realized along the positive real axis ($y > 0$), while quantum
real-time evolution corresponds to the unit circle $|y| = 1$.

For the thermal problem the variable $y$ goes to infinity as
temperature $T\to 0$, requiring us to consider the behavior of the map
at infinity.  To handle this, we extend the complex plane by including
the point at infinity, ${\widehat{\cal C}} = {\cal C}\cup\infty$,
which forms the Riemann sphere (Fig.  \ref{fig:basins}b).

\subsubsection{Loschmidt amplitude}
\label{sec:loschmidt-amplitude}

The central dynamical quantities of interest are the Loschmidt
amplitude $L(y)$, Eq. (\ref{eq:loschmidt_amp}), and the free energy,
$f(y)$, Eq. (\ref{eq:rate_function_def}).  The singularities of $f(y)$
encountered as $y$ traverses the unit circle (Fig. \ref{fig:basins})
signal the occurrences of dynamical quantum phase transitions during
the time evolution \cite{heylreview,heylprl}.

While one may, in principle, need to account for branch cuts and the
associated Riemann sheet structure, particularly due to the
multivalued nature of the complex logarithm, we note that this
ambiguity for log lies solely in the imaginary part. Since $f(y)$ is
real-valued and only its real part is physically relevant in the
present context, the complications related to multivaluedness can be
safely omitted for the purposes of this work.

\rgpicture

\section{Renormalization Group Framework}
\label{sec:rg}

The approach we adopt is the real-space Renormalization Group (RG)
\cite{pathria}.  The central principle is that the macroscopic
properties of a system are insensitive to the microscopic details of
its many degrees of freedom.  By systematically integrating out
short-scale degrees of freedom, one obtains an effective description
at larger scales in terms of renormalized interactions.  The defining
requirement of this transformation is that macroscopic quantities are
preserved.

To implement RG, the lattice is constructed hierarchically as shown in
Fig. \ref{fig:RGsteps}. The construction begins at generation $n=0$
with a single bond connecting two sites. At each subsequent
generation, every bond from the previous generation is replaced by two
new bonds connected in series. After $m$ recursive steps, the number
of bonds ($B_m$) and sites ($N_m$) are given by
\begin{equation}
    B_m = 2^m, \qquad N_m = B_m + 1.
    \label{eq:lattice_size}
\end{equation}
In the thermodynamic limit ($m \to \infty$), the number of sites and
bonds grow proportionally, and either can be used to define the system
size.  This iterative procedure generates a lattice that is equivalent
to a one-dimensional chain, and also shows a discrete scale invariance
(by construction), with a scaling factor of 2 at each generation (see
Fig. \ref{fig:RGsteps}).

The one-dimensional chain is decimated in a hierarchical manner that
mirrors the construction process mentioned above so that a chain of
$B_n$ bonds reduces to $B_{n-1}$ bonds, Eq. (\ref{eq:lattice_size}).
Details of the RG steps are given in Appendix~\ref{sec:rg-relations}
while a summary can be found in Fig. \ref{fig:RGsteps}. This
self-similar decimation preserves the discrete scale invariance and
leads to the RG transformations
\begin{subequations}
\begin{eqnarray}
  \label{eq:12}
  &&y' = R(y) \equiv   \frac{1}{2}\left( y + \frac{1}{y}\right),\hfill\\
  {\text{ and }} \qquad
  &&  Z_2(y) = c\, Z_{1}(y'), 
  {\text{  with }}  c=2y,  \label{eq:26}
\end{eqnarray}
\end{subequations}
where $y'$ is the renormalized $y$-parameter for the decimated chain,
and $c$ is the renormalization factor required to maintain the
partition function unchanged.

\subsection{Complex Dynamics of the RG Map}
\label{sec:complex}

The RG transformation, Eq. (\ref{eq:12}), is a rational function
$R(y)$ where numerator and denominator are polynomials with no common
factors. Iterating this map on the complex plane places the problem in
the framework of dynamics in one complex variable, or \emph{complex
  dynamics}.  Below we summarize a few relevant results of complex
dynamics \cite{milnor,beardon,carleson}.

Successive decimations reduce the number of degrees of freedom and
involve iterating the renormalization map $y'=R(y)$. The large-scale
behavior of the system is governed by the fixed points (fp) of this
map, defined by the condition $y^*=R(y^*)$.  The fixed points are
classified according to their multiplier, defined as
\begin{equation}
  \label{eq:15}
   \lambda = \left. \frac{d R(y)}{dy\,}\right|_{y^*},\ {\text{ for fp
     }} y^*,
\end{equation}
which determines their local stability properties. For the fixed point
at $y^*=\infty$, the multiplier is defined as
$\lambda=1/(dR/dy)|_{\infty}$, to account for the asymptotic behavior
of the map.

The classification is as follows \cite{comm1}:
\begin{enumerate}
\item repelling or unstable fixed point if $|\lambda|>1$;
\item attracting  or stable fixed point if $|\lambda|<1$;
\item superattracting fixed point if $|\lambda|=0$.
\end{enumerate}

For $R$, there are three fixed points, as given below with their
multiplicities (see Appendix \ref{sec:degree-rational-map}).
\begin{eqnarray}
  \label{eq:1}
  y_{\pm}^* &=& \pm1, {\text{ (stable), }} \lambda=0,\\
      y^*_u &=&\infty, {\text{ (unstable), }} \lambda=2. \label{eq:infi}
\end{eqnarray}
where the unstable fixed point represents the zero temperature, or,
equivalently, the ordered state of infinite coupling ($J\to\infty$).
On the Riemann sphere, Fig. \ref{fig:basins}, the unstable point is
the North pole, N.  The point $y_{+}^*=1$ represents the
infinite-temperature limit, where all spin configurations are equally
probable. This point is superattracting, reflecting the complete loss
of correlations and memory of microscopic details (the $J\to 0$
limit). This is the paramagnetic point.  The fixed point at $y = -1$
has no thermal interpretation, since it corresponds to a regime where
the Boltzmann weight is neither real nor positive. However, it remains
relevant for the quantum problem.  We show below that this fixed point
also corresponds to a paramagnetic phase.

\subsection{Flows and the Julia set}
\label{sec:flows-fixed-points}

Successive application of the renormalization group map produces the
sequence $y \mapsto R(y) \mapsto R(R(y)) \mapsto \dots$, which we
denote compactly as $y^{(j)} = R^{(j)}(y)$, where $R^{(j)}$ denotes
the $j$-fold composition of $R$.  The asymptotic behavior of this flow
is controlled by the fixed points of the map.

For the iterated map, the stable fixed points act as large-scale
attractors, and the set of initial values that flow into each
attractor defines its basin of attraction, known collectively as the
Fatou set. The boundary of the Fatou set in the extended complex plane
(Riemann sphere) is called the Julia set which contains the unstable
fixed points \cite{juliacomm,hastingslevi,Beardon1991,Peitgen2004}.  A
few general properties of the Julia set are listed in Appendix
\ref{sec:few-gener-prop}.  We also show below that for the RG map, the
Julia set ${\cal J}(R)$ determines the locations of the zeros of the
Loschmidt amplitude (partition function) in the complex-$y$ plane.

\subsubsection{Imaginary axis is the Julia set}
\label{sec:imaginary-axis-julia}

In the present case, the RG map has a pole at $y=0$. Under iteration
this point flows to $y=\infty,$ indicating that the pole flows
directly into the unstable fixed point at infinity (the North pole in
Fig.~\ref{fig:basins}b), which represents the ordered state of the
system. Since points that iterate into an unstable fixed point lie on
the Julia set, both $y=0$ and $y=\infty$ belong to ${\cal{J}}(R)$.
This means that the Julia set extends all the way to infinity on the
Riemann sphere.

This feature is closely tied to the physics of the one-dimensional
thermal Ising model. Because the model sits at its lower critical
dimension, thermal fluctuations are strong enough to destabilize the
ordered state at any finite temperature. The fact that the Julia set
reaches the unstable fixed point at infinity is a geometric expression
of this instability: the RG flow cannot sustain a stable ordered
phase, and trajectories are inevitably driven away from the ordered
point except at strictly zero temperature.

To determine explicitly the Julia set, we write $y = \xi + i\eta$ to
separate the flow into its real and imaginary parts.  Eq.
(\ref{eq:12}) yields
\begin{align}
    \xi' &= \xi \,\frac{\xi^2+\eta^2+1}{2(\xi^2+\eta^2)}, 
    \label{eq:xprime} \\
    \eta' &= \eta \,\frac{\xi^2+\eta^2-1}{2(\xi^2+\eta^2)}.
    \label{eq:etaprime}
\end{align}
Eq. \eqref{eq:xprime} shows that the sign of $\xi'$ is the same as
that of $\xi$. Hence, the RG flow never crosses the imaginary axis;
points with $\xi>0$ remain in the right half-plane and flow towards
the stable fixed point $y^*=1$, while those with $x<0$ remain in the
left half-plane and flow towards $y^*=-1$. The imaginary axis itself
is invariant under iteration but contains no finite-valued fixed
points. It is, therefore, {\em the Julia set} of the map, forming the
boundary between the two basins of attraction as shown in Fig.
\ref{fig:basins}.

To crosscheck that the extent of this Julia set is infinite, we
analyze the flow of points restricted to the imaginary axis.  Along
this line, the map for the imaginary component reduces to $\eta' =
(\eta^2 - 1)/(2\eta)$.  This expression shows that any point on the
imaginary axis that eventually flows to $\eta=\pm 1$ will, under
repeated iterations of the map, subsequently escape to infinity.
Consequently, the Julia set must be unbounded and, in particular,
extends along the entire imaginary axis. Appendix
\ref{sec:julia-set-rg} 
provides a different mathematical construction
of the Julia set, while Appendix \ref{sec:why-not-fractal} explains
why, in the present case, the resulting Julia set is not fractal.

\subsection{Free energy}
\label{sec:free-energy}

The decimation process shown in Fig. {\ref{fig:RGsteps}} reduces the
$n$th generation lattice containing $B_n$ bonds to the smaller
$(n-1)$th generation one having $B_{n-1}=B_n/2$ bonds (Eq.
\eqref{eq:lattice_size}).  The partition function of the $n$th
generation lattice, $Z_n(y)$, can be connected to that of the smaller
$(n-1)$th generation lattice by using the RG relations Eqs.
(\ref{eq:12}) and (\ref{eq:26}), as
\begin{equation}
  \label{eq:znbn}
    Z_n(y) = c(y)^{B_{n-1}} \, Z_{n-1}(y').
\end{equation}
We define a free-energy-like quantity  per bond 
\begin{equation}
f_n(y) = \frac{1}{B_n}\ln Z_n(y),
\label{eq:fndef}
\end{equation}
such that the thermodynamic limit satisfies
$$\lim_{n\to\infty} f_n(y) = -f(y),$$
with $f(y)$ defined in Eq. \eqref{eq:rate_function_def}.
Substituting Eq. \eqref{eq:znbn} into Eq. \eqref{eq:fndef} gives the recursion
\begin{equation}
  f_n(y) = \frac{1}{2}\, f_{n-1}(y') + \frac{1}{2}\, g(y),
  \textrm{ with } g(y) = \ln c(y).
  \label{eq:frecur}
\end{equation}
This equation relates the free-energy densities across successive
generations.

Iterating the RG map, we define
\begin{equation}
  y^{(j)} \equiv R^{(j)}(y)= 
           \underbrace{R(R(\dots R}_{j \text{ times}}(y))), 
            \quad{\text{with }}\, y^{(0)} = y, 
            \label{eq:iterates}
\end{equation}
so that $y^{(j)}$ denotes the renormalized coupling after $j$
decimation steps.  Eq. \eqref{eq:frecur} then leads to the exact
series
\begin{equation}
  f_n(y) = \sum_{j=0}^{n-1} \frac{1}{2^{j+1}}\, g\!\left(y^{(j)}\right) 
               + \frac{1}{2^n}\ln Z_1\!\left(y^{(n)}\right).
               \label{eq:frate}
\end{equation}
The summation on the right hand side collects the contributions from
each RG generation, with the coupling at that stage given by
$y^{(j)}$, as illustrated in Fig.~\ref{fig:RGsteps} where $y', y''$
are used for $y^{(1)}, y^{(2)}$, respectively.  All dependence on the
boundary conditions is confined to the final term involving $Z_{1}$.
The expression in Eq.~\eqref{eq:frate} converges rapidly and may be
evaluated for large $n$ to obtain the limiting function $f(y)$.

\subsection{Julia set and the zeros of the  partition function}
\label{sec:julia-set-zeros}

The Renormalization Group (RG) provides a method for determining the
zeros of the partition function in the thermodynamic limit.

A zero of the partition function $Z_n(y)$ is a value of the parameter,
$\zeta_j$, for which $Z_n(\zeta_j) = 0$. Applying this condition to
the recurrence relation Eq. \eqref{eq:znbn} implies that
$Z_{n-1}(R(\zeta_j))$ must be zero. This means that the value
$R(\zeta_j)$ must itself be a zero of the $(n-1)$th system's partition
function. We can call such a zero $\zeta'_k$. This leads to the
fundamental relationship:
\begin{equation}
  R(\zeta_j) = \zeta'_k
\end{equation}

This equation dictates that the zeros of the large lattice ($\zeta_j$)
are the \textit{preimages} of the zeros of the smaller lattice
($\zeta'_k$). In practice, this equation is used to find the unknown
zeros of a larger system by starting with the known zeros of the
smallest system (see Fig. \ref{fig:triangle}) and solving for their
preimages \cite{luck}. This \textit{backwards} iteration generates the
full set of zeros for any generation.

In the thermodynamic limit, the set of all these zeros converges to
the \textit{Julia set} of the map $R(y)$. This occurs because the
process of finding preimages is an iteration of the inverse RG map.
This backward flow naturally takes points away from attractive fixed
points and towards the chaotic boundary that separates their basins of
attraction, which is the Julia set.

\trianglefig

\subsubsection{Zeros of the 1-d chain}
\label{sec:zeros-1-d}

The general principle applies to both open and periodic boundary
conditions, but the outcomes are dramatically different due to their
initial seed of zeros:

\begin{itemize}
\item \textit{For the open chain}, the process starts with a single
  zero at $y=-1$, as $Z_1= 2 (y+1)$ (see Fig. \ref{fig:triangle}).
  This is an exceptional case because this zero is also a
  \textit{stable fixed point of the RG map} (i.e., $R(-1)=-1$).
  Therefore, the preimages of the zero are simply mapped back onto
  itself. Consequently, all zeros for all generations remain at the
  isolated point $y=-1$ and do not spread out to form the Julia set.
    
\item \textit{For the periodic chain}, the process starts with initial
  zeros that are \textit{not} fixed points, e.g., if we start with a
  triangle of three spins (see Fig. \ref{fig:triangle}). Finding the
  preimages of these points generates a new and more complex set of
  zeros.  Repeating this backward iteration spreads the zeros out, and
  in the infinite-system limit, they become dense on the Julia set,
  leading to a rich structure of DQPTs.
\end{itemize}
We see that, unlike in standard thermodynamic settings, boundary
conditions, open versus closed chains, significantly affect the
behavior of the system.  The difference is also seen in the free
energy $f(y)$, as shown in Fig. \ref{fig:open-vs-periodic}. These
effects are examined in detail in subsequent sections.

\subsubsection{Direct determination of zeros}
\label{sec:direct-determ-zeros}

The identification of the imaginary axis with the Julia set is
confirmed by analyzing the zeros of $L(y)$. For a finite {\it periodic
  chain} of length $N$, the zeros of the partition function are given
exactly by
\begin{equation}
    y_k = -i \cot\left(\frac{(2k+1)\pi}{2N}\right), 
    \qquad k=0, 1,\dots,N-1,
    \label{eq:zeros}
\end{equation}
all of which lie on the imaginary axis. The zeros are obtained by a
transfer matrix approach as discussed in Appendix
\ref{sec:density-zeros} and Sec.  \ref{sec:single-bond-case}.

As $N \to \infty$, these zeros can be described by a density
\begin{equation}
  \label{eq:13}
  \rho(\eta)= \frac{1}{\pi} \frac{1}{1+\eta^2} \text{ (per spin),}
\end{equation}
where $y=i\eta$ in the complex-$y$ plane and 
$\eta\in (-\infty,+\infty)$. 
Details are in Appendix \ref{sec:density-zeros}.

The density at $y=\pm i$ on the unit circle is $\rho(1)=1/2\pi$. The
usual electrostatic analogy ensures that every intersection of the
unit circle with the imaginary axis would produce a continuous $f(y)$
but with a slope discontinuity determined by $\rho(1)$.  This confirms
that the Julia set governs the onset of dynamical singularities.

\subsubsection{Julia set and DQPT}
\label{sec:julia-set-dqpt}

To summarize the key ideas: The quantum evolution corresponds to
motion along the unit circle $|y|=1$ in the complex-$y$ plane.  At
each instant, this point serves as the initial condition for the RG
map. By iterating the RG map, we coarse-grain the system in space,
revealing its dominant behavior at a macroscopic scale. The fixed
point this flow leads to determines the phase of the system at that
specific instant.

Further, our RG transformation possesses a characteristic Julia set.
These are the points that do not flow to stable fixed points, and they
are also the locations of the zeros of the partition function.

On  combining these facts, we arrive at our central claim: 
\begin{quote} 
  DQPTs correspond to the intersection of the unit circle with the
  Julia set. These intersection points are precisely the critical
  times where the system undergoes a dynamical quantum phase
  transition.
\end{quote}

For TFIM on a periodic chain, the DQPTs are first-order transitions at
$y=\pm i$, or respectively, at times
\begin{subequations}
\begin{equation}
  \label{eq:38}
  t_{c1}=  \frac{\pi}{2} \frac{\hbar}{J}, \quad  t_{c2}=  \frac{3\pi}{2} \frac{\hbar}{J}
\end{equation}
and then periodically at
\begin{equation}
  \label{eq:16}
  t_{cn}=  (2n-1)\frac{\pi}{2} \frac{\hbar}{J},\quad n=3,4,5... ,
\end{equation}
\end{subequations}
where we have put back $\hbar$ to put the time scales in proper
dimensional form.

\subsubsection{Basins of attraction and DQPT}
\label{sec:basins-attr-dqpt}

As established in Sec.~\ref{sec:flows-fixed-points}, and
\ref{sec:julia-set-zeros}, the Julia set coincides with the imaginary
axis, separating the basins of attraction of $y^*=1$ and $y^*=-1$.
Starting from $y=e^{iJt}$ on the unit circle, we follow its RG
trajectory to identify the fixed point it is attracted to. See Fig.
\ref{fig:basins}.

\dqptplot

\begin{itemize}
    \item For $0<Jt<\pi/2$, the system lies in the basin of $y^*=1$
      (equivalent to the infinite temperature paramagnetic phase).
    
    \item At $Jt=\pi/2$, the unit circle intersects the Julia set. The
      RG flow goes eventually to infinity (equivalent to the zero
      temperature ordered state).  
      The free energy $f(t)$ develops nonanalyticity at this time.  

      In the standard notation of critical phenomena \cite{pathria},
      $f(y) \sim \abs{y-y_c}^{2 - \alpha}$ near the transition at $y =
      y_c$. The exponent $\alpha$ is related to $\lambda,$ Eq.
      (\ref{eq:infi}),  by $\alpha = \frac{\ln \lambda}{\ln 2}$, for
      the scale factor $=2$ (Fig. \ref{fig:RGsteps}). Since
      $\lambda=2$ for the unstable fixed point, $\alpha =1$. This
      shows the linear behavior of $f(t)$ near $t_c$ indicating a
      slope discontinuity at the transition point, as discussed in the
      last subsection.

    \item For $\pi/2<Jt<3\pi/2$, the flows are in the basin of
      $y^*=-1$, corresponding to a  paramagnetic phase (no
      thermal analog). We call this an intermediate phase.

    \item At $Jt=3\pi/2$, the unit circle intersects the Julia set
      again, producing another DQPT. This is similar to the
      $Jt=\pi/2$ case.

    \item For $3\pi/2<Jt<2\pi$, the system again  enters  the basin of
      $y^*=1$.

\end{itemize} 

Thus, time evolution generates oscillatory transitions between
distinct phases, with each DQPT arising precisely from the
intersection of the unit circle with the Julia set.

The above discussion when combined with Eq. (\ref{eq:frate}) leads to
the plot of the free energy $f(t)$ versus time for a periodic chain as
shown in Fig. \ref{fig:open-vs-periodic}.  The curve for the periodic
chain shows cusps at the critical times $Jt=\pi/2$ and $Jt=3\pi/2$,
exactly where the unit circle intersects the Julia set. These are the
DQPTs.

There is however a surprising suppression of these phase transitions
by changing the boundary conditions, as seen in the free energy curve
(Fig.  \ref{fig:open-vs-periodic}) for the open chain.  Here, one
instead observes orthogonal catastrophe at $Jt=\pi$ in the form of a
log-divergence of $f(t)$. This signifies the evolution of the system to
a state orthogonal to its initial state, $\ket{\psi_0}$, Eq.
(\ref{eq:14}).  

Such a strong sensitivity of the bulk behavior to the boundary
conditions is not expected in conventional thermal transitions.  This
difference is discussed in the next section.

\section{Intermediate Phase and Boundary Conditions}
\label{sec:interm-phase-bound}

\ringopen

The fundamental difference between open and periodic boundary
conditions is topological because an open chain cannot be continuously
transformed into a ring. To investigate how this topology affects the
system's physical properties, we induce a crossover by modifying the
Hamiltonian by tuning a single bond.  Gradually turning off the
coupling of a single bond ($J_{\text{b}}$) connecting the chain ends,
changing the system from a periodic ring ($J_{\text{b}}=J$) to an open
chain ($J_{\text{b}}=0$).

As a complementary approach, we utilize the Transfer Matrix method to
study this crossover in detail.

\subsection{Single bond boundary}
\label{sec:single-bond-case}

Consider a ring of $N$ sites with periodic boundary conditions, where
all bonds initially have the same coupling strength $J$. To probe the
effect of changing the topology, one bond (say, between spin 1 and
spin $N$, Fig. \ref{fig:ring}) is modified to have a different
coupling $\jb$. As $\jb\to0$, the spins connected by that bond become
effectively decoupled, turning the ring into an open chain with open
boundary conditions.  This particular coupling, $J_b$, will be called
the boundary coupling or boundary bond.

Phase transitions strictly occur only in the thermodynamic limit; for
finite systems, smooth peaks appear instead of genuine singularities.
Therefore, the topological transformation involves taking two limits,
namely
\begin{equation}
  \label{eq:4}
  \lim \jb\to 0, \quad {\text{and }} \lim N\to\infty,
\end{equation}
and the order of these limits matters.  If the thermodynamic limit
$N\to\infty$ is taken first, the system retains its bulk properties
regardless of the value of $\jb\neq 0$. In contrast, taking $\jb\to0$
first introduces boundary effects that may persist even in the
thermodynamic limit.

The partition function can be obtained by using the transfer matrices
\cite{pathria} and as discussed in Appendix A, it is given by
\begin{equation}
  \label{eq:6}
  Z_N= \frac{1}{2^N}\left[\lambda_1^{N-1} (\yb + 1) +
  \lambda_2^{N-1}  (\yb - 1)\right ],
\end{equation}
 where $y_{\text{b}}=e^{iJ_{\text{b}}t}$ is the boundary bond weight, and 
\begin{equation}
  \label{eq:17}
 \lambda_1=y+1, {\text{ and }} \lambda_2=y-1, 
\end{equation}
are the eigen values of the bulk transfer matrix. The zeros of the  partition function
as discussed in Sec. \ref{sec:direct-determ-zeros} follow from
Eq. (\ref{eq:6})  with $\jb=J$.

For $N\to\infty$, DQPTs occur at times when the leading and subleading
eigenvalues become degenerate in magnitude, i.e., when
  $|\lambda_1/\lambda_2|=1.$ 
For $\jb=J$, the transitions occur at times
$t_c$ defined in Eq. (\ref{eq:16}), and the free energy density for
$N\to\infty$ takes the piecewise form determined by the larger of the
two eigenvalues, $\max(|\lambda_1|, |\lambda_2|)$.  Using Eqs.
(\ref{eq:rate_function_def}) and (\ref{eq:17}), we obtain
\begin{eqnarray}
  \label{eq:18}
  f(t)= \left\{ \begin{array}{lll}
                -\ln |\lambda_1| = & -\frac{1}{2}\ln \cos^2 \frac{Jt}{2},& {\text{for }}0\le t\le \frac{\pi}{2J},\\[6pt]
                -\ln |\lambda_2| = &  -\frac{1}{2}\ln \sin^2 \frac{Jt}{2},& \text{for }
                 \frac{\pi}{2J}\le t\le \frac{3\pi}{2J},
                  \end{array}\right.                  
\end{eqnarray}
with periodicity of $2\pi/J$. These two branches, which characterize
the distinct phases, are shown in Fig. \ref{fig:open-vs-periodic}.
The phase described by $\lambda_2$ corresponding to the fixed point
$y=-1$ (Eq.~\eqref{eq:1}) is to be called the \textit{intermediate
  phase} (no thermal analogue).  The two free energies are related by
symmetry because $|\lambda_1(t)|=|\lambda_2(t+\frac{\pi}{J})|$,
indicating that the intermediate phase is also a paramagnetic phase.

However, for finite $N$, the nonanalytic behavior is smoothened, and
the free energy becomes analytic for all times due to finite-size
effects.  Nevertheless, a finite-size precursor to the transition
persists as a peak in the free energy. As $f(t)$ crosses over from a
$\lambda_1$-dominated regime to a $\lambda_2$-dominated one, the peak
occurs when the two terms of the partition function are comparable.
Therefore the peak occurs when
\begin{equation}
  \label{eq:8}
  \left | \frac{\lambda_1}{\lambda_2}\right|= 
    \left(\left| \frac{\yb-1}{\yb+1}\right|\right)^{1/N}.
\end{equation}
This defines the peak position parametrically in time, as shown in
Fig. \ref{fig:TM_012}a.  In the special case $\yb=1$, corresponding to
an open chain, Eq.\eqref{eq:8} reduces to $\lambda_1=0,$ or
equivalently $y\equiv e^{iJt}=-1$ as one would expect ($Jt=\pi$).

\tmzeroonetwo

Figure \ref{fig:TM_012}a illustrates the $\jb$--dependence of the free
energy $f_N(t)$ computed via the transfer matrix method for a finite
ring of size $N=100$.  As $\jb$ decreases, the peaks in the free
energy move closer together and gradually approach the divergent
behavior of the open chain at $Jt=\pi$. Conversely, for any fixed
$\jb>0,$ the $N\to\infty$ limit restores the DQPT curve associated
with $\jb=J$, as shown in Fig. \ref{fig:TM_012}b regardless of how
small $\jb$ is.

To verify that the curves seen for $\jb\neq 0$ correspond to the same
intermediate phase (described by $\lambda_2$), we isolate the bulk
contribution by subtracting the boundary term and defining a scaled
free energy
\begin{equation}
  \label{eq:2}
  {\overline{f}}(y)=f(y) - \dfrac{1}{N} \text{Re} \ln{(\yb-1)},
\end{equation}
As shown in Fig \ref{fig:TM_012}c, all the resulting curves collapse
onto the characteristic rate function for the intermediate phase.

This demonstrates that the intermediate phase is suppressed in the
open-chain limit due to boundary effects. For any fixed $N$, one can
find a range of time such that the boundary contribution, ${\text{Re}}
\ln (\yb -1)$, becomes comparable to the bulk term.  In other words,
there will be a range of time around $J t=\pi$ such that
\begin{equation}
  \label{eq:9}
  \ln \sin^2 \frac{\jb t}{2} \sim N f,
\end{equation}
which pushes the peak position towards this point of time.

\subsubsection{Speed limits and pbc}
\label{sec:speed-limits-pbc}

The influence of the boundary bond can be understood by comparing
relevant quantum time scales \cite{frey,qsl}. For a bond of strength
$\jb$, the quantum speed limit (QSL) for state evolution across the
bond is given by the Mandelstam--Tamm or Margolus--Levitin bounds
\cite{kaur1}, yielding an estimate 
  $\tau_{\text{qsl}}\sim \pi \hbar/\jb.$ 
This sets the minimum time required for significant
dynamical changes across the boundary. In contrast, the time for
information to propagate from one side of the chain to the other,
assuming a local perturbation at one end, can be estimated using the
Lieb--Robinson bound \cite{liebrob,hastings}. For a chain of length
$Na,$ where $a$ is the lattice spacing, and with a Lieb-Robinson
velocity $v_{LR}\sim J a,$ the characteristic propagation time is
$t_{\text{LR}} \sim N/J$.  The boundary bond becomes dynamically
irrelevant when 
  $\tau_{\text{qsl}}\gg t_{\text{LR}},$ 
which translates to the condition $N\jb/J \ll 1$.

For a finite chain, as $\jb \to 0$, the peaks in $f(t)$ signaling
DQPTs, originally occurring at the critical times $t_{c1}$ and
$t_{c2}$ (Eq.~\eqref{eq:38}), become progressively rounded and
shifted, diminishing the influence of the intermediate phase. In the
limit $\jb = 0$, these two peaks coalesce into a single one at $t =
\frac{\pi \hbar}{J}$, whereby the pair of DQPTs is effectively
replaced by an orthogonality catastrophe.

\section{Conclusion}
\label{sec:conclusion}

In this work, we established a connection between the non-equilibrium
phenomena of dynamical quantum phase transitions (DQPTs) and the
mathematical framework of complex dynamics through a real-space RG
analysis of the one dimensional transverse-field Ising Model. Our
central finding is that DQPTs, which manifest as non-analyticities in
the Loschmidt echo, occur precisely at the moments in time when the
system's unitary evolution path intersects the Julia set of its
corresponding RG transformation. It is important to note that the
system's evolution is purely unitary, governed by a timescale
proportional to $\hbar/J, J$ being the interaction coupling constant.
Consequently, these phenomena are inherently quantum mechanical,
lacking a direct classical analog in the $\hbar \to 0$ limit. By
mapping the quench dynamics to an RG flow in the complex plane of the
parameter $y=e^{iJt}$, we identified the imaginary axis as the Julia
set, which is the boundary separating the basins of attraction for the
stable fixed points at $y=\pm 1$. For a periodic chain, the time
evolution, represented by a trajectory along the unit circle,
periodically crosses this Julia set, giving rise to a series of DQPTs
at critical times $t_{cn} = (2n+1)\pi \hbar/2J$.  In contrast to the
periodic chain, it is found that an open chain exhibits a complete
suppression of DQPTs. By tuning a single boundary bond, the ring
topology is effectively broken, resulting in the merger of distinct
transitions into a single logarithmic divergence. This is indicated by
an orthogonality catastrophe at $t=\pi \hbar/J$. This effect is
rationalized using quantum speed limits, which impose constraints on
the dynamical evolution along two distinct geodesic paths on the ring.
Our results highlight the role of topology and boundary conditions
together with interactions (here spin-spin interaction) in shaping the
real-time dynamical phase structure of quantum many-body systems.

Real-space renormalization group (RG) methods are widely applicable
across systems, models, and dimensionalities when studied across
scales, and RG transformations are expected to be analytic functions
of real Boltzmann weights. Our approach thus provides a distinct
perspective on quantum quenches of various quantum models. Because
Julia sets are highly sensitive to the RG map, the method we presented
can serve as a benchmark for different approximate renormalization
schemes. Future work could expand this approach to include other
quench protocols beyond those discussed here.

\appendix

\section{The RG relations}
\label{sec:rg-relations}
To derive the RG equations summarized in Fig. \ref{fig:RGsteps}, we
use the transfer-matrix formulation of the 1-dimensional Ising model
(see, e.g., Ref. \cite{pathria}).

\subsection{Transfer matrix and Partition functions}
\label{sec:part-funct-transf}

For the 1-D Ising chain, the partition function can be constructed
iteratively by introducing a transfer matrix ${\bm{T}}$ that encodes
both the Boltzmann weight associated with the interaction between
neighboring spins and any combinatorial factors that arise as one spin
is appended to a chain of length $n$.  Denoting the two possible spin
states by $\ket{+}$ and $\ket{-}$, each state of the $(n+1)$th spin is
compatible with either of the two states of the $n$-th spin.  These
four possibilities are incorporated through a $2\times 2$ transfer
matrix which takes the form
\begin{equation}
  \label{eq:29}
 \bm{T}=\left(\begin{array}{ll}
                                  y& 1\\[2pt]
                                  1& y
                                  \end{array}
                                  \right),
\end{equation}
with eigenvalues 
\begin{equation}
  \label{eq:39}
  \lambda_1 = y+1, \quad {\text{ and }} \quad \lambda_2 = y-1.
\end{equation}
so that, for fixed end spins $s_1,s_N=\pm$, the restricted partition
function is given by
\begin{equation}
  \label{eq:31}
  Z_N(s_1,s_N) = \bra{s_1} {\bm{T}}^{N-1}\ket{s_N}.\,
\end{equation}
corresponding to the appropriate matrix element of the matrix.

\subsubsection{Special Partition functions}
\label{sec:spec-part-funct}
Two choices of boundary conditions are considered in this paper.

(a) Open boundary conditions where $s_1$ and $s_N$ can have
unrestricted possibilities so that the partition function is given by
\begin{equation}
  \label{eq:36}
  Z_{N|{\text{free}}}=\sum_{\substack{s_1=\pm\\ s_N=\pm}} Z_N(s_1,s_N).
\end{equation}

(b) Periodic boundary conditions (pbc) for which an additional
interaction couples spins on sites $N$ and $1$ (see Fig.
\ref{fig:ring}. The partition function is then the sum of the diagonal
elements,
\begin{equation}
  \label{eq:37}
   Z_{N|{\text{pbc}}}= \Tr \bm{T}^N = \lambda_1^N + \lambda_2^N.
\end{equation}

(c) If one link in case (b) is given a different weight 
$\yb = e^{i\jb t}$, then the transfer matrix for that bond is 
${\bm{T}}_b$ which is similar to ${\bm{T}}$ of Eq. (\ref{eq:29})
except for the replacement of $y$ by $\yb$,
\begin{equation}
   \label{eq:a7}
  \bm{T}_{\text{b}}=\left(\begin{array}{ll}
                                  \yb& 1\\
                                  1& \yb
                                  \end{array}
                                  \right),
                  \end{equation}
with eigenvalues $\lambda_{1b}=\yb+1, \lambda_{2b}=\yb-1.$  
The modified partition function is given by
\begin{eqnarray}
  \label{eq:40}
  Z_{N|{\text{mod}}} &=& \frac{1}{2^N} \Tr \bm{T}^{N-1}\bm{T}_b
  \nonumber\\ 
                    &=&   \frac{1}{2^N}\left[\lambda_1^N (\yb + 1) +
  \lambda_2^N  (\yb - 1)\right ].
\end{eqnarray}

\subsection{Decimation and RG relations}
\label{sec:decim-rg-relat}

Decimating every second spin corresponds to summing over the middle
spin of a three-spin block. With boundary spins as $\alpha,\beta=\pm$,
as in Fig.  \ref{fig:RGsteps}, the partition function for a three-spin
block is $Z_1(y|\alpha\beta)=\bra{\alpha}{\bm{T}}^2\ket{\beta}.$
Therefore, the effective interaction between spins 1 and 3 after
decimation is encoded in a renormalized transfer matrix ${\bm{T}}'$,
which must have the same form as ${\bm{T}}$ but with a new coupling
parameter $y'$.

Since ${\bm{T}}^2$ generates the same two-spin partition function (up
to an overall normalization), we require
\begin{equation}
  \label{eq:32}
  {\bm{T}}^2 =  c\, {\bm{T}}',
\end{equation}
where $c$ is a normalization factor ensuring that $\bm{T}'$ has the
same structural form as $\bm{T}$, i.e.,
\begin{equation}
  \label{eq:35}
  \bm{T}' = \left(\begin{array}{ll}
                                  y'& 1\\
                                  1& y'
                                  \end{array}
                                  \right).  
\end{equation}

Direct evaluation gives 
\begin{equation}
\label{eq:34}
    \bm{T}^2 = \left(\begin{array}{ll}
                                  y^2+1& 2y\\
                                  2y& y^2+1
                                  \end{array}
                                  \right),
\end{equation}
which, on equating  with Eq.~\eqref{eq:35}, yields
\begin{equation}
    c y' = y^2 + 1, 
    \qquad 
    c = 2y,
\end{equation}
and
\begin{equation}
    y' = \frac{y^2 + 1}{2y}.
\end{equation}
These are the RG equations, Eqs.~(\ref{eq:12}) and (\ref{eq:26},) in
the text.  The diagonal and the off-diagonal terms of Eq.
(\ref{eq:34}) are the same-spins and different-spins partition
functions in Fig. \ref{fig:RGsteps}.

\section{Degree of a  rational map and the number of fixed points}
\label{sec:degree-rational-map}
The degree of a rational map $R(y)$ is defined as the maximum of the
degrees of the polynomials in its numerator and denominator.  A
rational map of degree $d$ has $d+1$ fixed points on the Riemann
sphere.  If
\begin{equation}
  \label{eq:27}
  R(y)\to \infty \quad {\text{as}} \quad  y\to \infty,   
\end{equation}
then $y=\infty$ is one of the fixed points. The remaining fixed points
follow from the finite-$y$ condition $y=R(y)$, which reduces to a
polynomial equation of degree $d$.

For our one-dimensional problem, the RG map has degree
${\text{deg}}(R)=2$, and therefore possesses three fixed points in
total. Because $R(y)$ satisfies Eq.~(\ref{eq:27}), one fixed point
lies at $y = \infty$, while the other two are obtained from the
quadratic equation $y=R(y)$. These finite-valued fixed points are
given in Eqs.~(\ref{eq:1}) and (\ref{eq:infi}).
\section{A few general properties of the Julia set}
\label{sec:few-gener-prop}

The Julia set, ${\cal J}(R)$, associated with a rational map $R$
possesses several important general properties as below.
\begin{enumerate}
\item It has no isolated points.
\item It is completely invariant under $R$, i.e, $R({\cal J}(F)) =
  {\cal J}(F)$.
\item It contains all unstable fixed points of the map.
\item It contains infinitly many points (in fact, is dense in itself)
  and is typically a fractal for non-integrable maps. 
\end{enumerate}
For the RG map considered here, the trajectories can be written in
closed form through a sequence of transformations
(Appendix~\ref{sec:why-not-fractal}). As a result, the Julia set is
not fractal but reduces to a smooth curve, specifically the imaginary
axis.

\section{Julia set of the RG map }
\label{sec:julia-set-rg}

\subsection{Construction of the Julia set}
\label{sec:constr-julia-set}

\begin{subequations}
Consider the RG map of Eq.~(\ref{eq:12})
\begin{equation}
  \label{eq:24}
  R(y) = \frac{1}{2} \!\left(y + \frac{1}{y}\right),
\end{equation}
defined on the extended complex plane $\widehat{\mathbb{C}}$ (the
Riemann sphere, Fig.~\ref{fig:basins}b).  We show that its Julia set
is the imaginary axis by following the steps below.

\begin{enumerate}
  \item 
The M\"obius transformation $M$
    \begin{equation}
      \label{eq:19}
      z = M(y) \equiv  \frac{1 - i y}{1 + i y},
    \end{equation}
    yields the map conjugate to $R$ as
    \begin{equation}
      \label{eq:20}
      z' = R_M(z) \equiv z^2.
    \end{equation}

  \item 
The Julia set of the squaring map $R_M$ is the unit circle,
    \begin{equation}
      \label{eq:21}
      {\cal{J}}(R_M) = S^1 = \{\, |z| = 1 \,\}.        
    \end{equation}
    Indeed, under repeated iteration of $R_M$, all points with $|z|<1$
    converge to $z=0$, while those with $|z|>1$ diverge to infinity.
    Points on $|z|=1$, $z=e^{i\theta}$, remain on the unit circle for all iterations.

  \item 
Since M\"obius conjugacy preserves the Julia set structure
    \cite{milnor,beardon,carleson},
    \begin{equation}
      \label{eq:22}
      {\cal{J}}(R) = M^{-1}(S^1),
      \, 
      \text{with }
      M^{-1}(z) = -i\,\frac{1 - z}{1 + z}.
    \end{equation}
    For $z = e^{i\theta} \in S^1$, with the polar angle $\theta\in[0, 2\pi)$,
    \begin{equation}
      M^{-1}(e^{i\theta}) 
      = -i\,\frac{1 - e^{i\theta}}{1 + e^{i\theta}}
      = i\,\tan\!\left(\frac{\theta}{2}\right),
    \end{equation}
    which is purely imaginary. As
    $-\infty<\tan\!\left(\frac{\theta}{2}\right)<\infty$, it follows that
    \begin{equation}
      \label{eq:23}
      {\cal{J}}(R) = i\;\mathbb{R},
    \end{equation}
    i.e., the Julia set of the RG map~(\ref{eq:24}) is the imaginary axis.
\end{enumerate}
\end{subequations}

\subsection{Why not a fractal?}
\label{sec:why-not-fractal}

The map $R(y)$, which can be transformed to the squaring map
(Eq.~\eqref{eq:20}), produces a smooth Julia set, unlike maps such as
$z' = z^2 - 1$, whose Julia set exhibit an intricate fractal
structure.

As discussed in Refs.~\cite{milnor,beardon,carleson}, the geometry of
the Julia set ${\cal{J}}(R)$ is determined almost entirely by the
behavior of the critical orbits of $R$. Critical points are the zeros
of $dR(y)/dy$ or the poles where $R$ fails to be locally invertible.
The forward images $R(y_c),\; R(R(y_c)),\; \ldots,$ of a critical
point $y_c$ form its \emph{critical orbit}. The union of all such
orbits, often called the \emph{postcritical set}, governs the global
nonlinearity of the map.

For the map $R(y)$ of Eq.~(\eqref{eq:24}), the critical points are
$y_c = \pm 1$, which are themselves fixed points of $R$. Their
critical orbits are therefore trivial, since there is no motion under
iteration. In such cases the map is \emph{postcritically finite}, i.e.
there is no infinite critical orbit, and the Julia set ${\cal{J}}(R)$
is a smooth one-dimensional manifold, typically a circle, line, or
smooth algebraic curve. The dynamics is \emph{integrable}, with
iterates expressible in closed form, as in the present case through
the M\"obius transformation of Eq.~(\eqref{eq:19}).  \\[24pt]

\section{Density of zeros}
\label{sec:density-zeros}

We determine the large-$N$ density of zeros of the partition function,
Eq. (\eqref{eq:13}), $Z_N=(y-1)^N+(y+1)^N,$ along the imaginary axis
of the complex $y$-plane.

\begin{subequations}
The zeros  are given by
Eq. (\eqref{eq:zeros}), $y_k= - i \eta_k$ where
\begin{equation}
  \label{eq:28}
 \eta_k = \cot\left( \frac{(2k + 1)\pi}{2N} \right), \quad k = 0, 1, 2, \ldots 
\end{equation}
As $N \to \infty$, the argument of the cotangent function becomes
dense in the interval $(0, \pi),$ and since $\cot(\theta)$ is a
strictly decreasing function mapping $ (0, \pi) \to \mathbb{R},$ the
set $\{ \eta_k \}$ becomes dense in $ \mathbb{R},$ allowing us to
define a density of zeros.

To compute the density of zeros $ \rho(\eta)$ per spin, in the large
$N$ limit, we first note that the angular spacing between successive
zeros is $\Delta \phi = \frac{\pi}{N},$ so that 
 $\rho_\phi(\phi) = \frac{1}{N}\;\frac{N}{\pi}.$ 
With a change of variable from $ \theta $ to $ \eta = \cot(\phi),$ 
the density transforms as
\begin{equation}
  \label{eq:30}
 \rho(\eta) = \rho_\phi(\phi) \left| \frac{d\phi}{d\eta} \right|.
\end{equation}
Now, substituting  $\frac{d\phi}{d\eta} = - \sin^2(\phi) = - (1 + \eta^2)^{-1}$,
we obtain the large-$N$  density of zeros along the imaginary axis as
\begin{equation}
  \label{eq:33}
  \rho(\eta) = \frac{1}{\pi} \frac{1}{1 + \eta^2}.  
\end{equation}
\end{subequations}
This is a scaled Cauchy (Lorentzian) distribution over
$-\infty<\eta<\infty,$ with 
$\int_{-\infty}^{\infty} \rho(\eta) d\eta = 1.$



\begin{thebibliography}{99}
\bibitem{fidel} As such, $L(t)$ serves as a sensitive measure of
  information loss and the dynamical stability of a quantum states
  under unitary evolution. In a general context, $|L|^2$ is also
  called the fidelity.
\bibitem{comm1} Other types of fixed points exist (e.g., neutral or
  periodic), but only those relevant for the present work are noted
  here.
\end{thebibliography}
\end{document}